# A Flickering Resonance Degeneracy on Entropy-Ruled Charge Transport in the Extended Hopping Sites for Biological Systems: Role of Static-to-Dynamic Disorder Transition

K. Navamani[1,(a], M. Pavalamuthu[1]
Department of Physics
KPR Institute of Engineering and Technology
Coimbatore-641 407, India.

[a)]Author to whom correspondence should be addressed: pranavam5q@gmail.com

## ABSTRACT

Charge transport (CT) in biological systems is of great interest due to its role in various functional activities like photobiology, bioenergetics, redox catalytic and other metabolic activities, etc. It has been observed by various studies that the presence of a dynamic disorder in biomolecules facilitates charge dynamics in the intermediate transport regime, i.e., far from hopping and towards a bank-like mechanism. Hitherto, the dynamic disorder (in a time scale, coupling between electronic and nuclear dynamics) weightage on CT in molecules is not well-established for the measurement of molecular conductivity from small to long-range ordering. With this motivation, we propose the flickering resonance-coupled entropy-ruled charge transport theory for transfer-rate and diffusion-based mobility (D/μ) calculations. The proposed analytical formalism incorporates the impact of dynamic disorder-correlated degeneracy (in a flickering-resonance manner) on the entropy-ruled electron transfer rate and diffusion-mobility, which are valid for localized hopping, delocalized band transport, and regimes in between. By this approach, it has been observed that the dynamics-driven electronic site matching probability enhances the degeneracy, which is quantified by the differential entropy. The analysis clearly shows that for ultra-fast dynamical and degenerate cases, the entropy-ruled mobility is transformed as band mobility, rather than the thermally activated hopping or diffusion process. During biochemical reactions (like enzymatic reactions, the formation of oxidizing/reduction equivalents (holes/electrons) are transported through the redox-active centres/residues, which helps to protect the protein from oxidative stress. In this regard, herein the developed transfer rate theory of an extended hopping sites system is tested and validated for a few limited sites to extended dynamical model systems under different sets of electric field-assisted site energy flux conditions.

## METHODS

The biological charge transfer mechanism plays a central role in electron and proton transfer, or any ionic-mediated all biochemical reactions, and importantly acts as a fundamental key descriptor for ionic diffusion-based conductivity, such as $Na^{+}$, $Ca^{2+}$, $Mg^{2+}$, $K^{+}$, and $Cl^{-}$, etc.[1-3] The ionic channelized motion usually regulates the electrolyte balance, pH levels, protein protection from redox stress, cellular mechanisms, and normalizes the various enzymatic reactions and metabolic activities in our bodies. The effect of structural fluctuations (e.g., protein folding dynamics) has a significant impact on CT, and it leads to a crossover from hopping to band-like carrier motion.[1,3] Here, the energetically relaxed electron (polaron) dynamics in the distorted biological medium are yet to be well-explored. As observed from various studies, the dynamic disorder minimizes the energetic barrier, and hence it leads to ballistic charge motion rather than localized transport along the ionic channel.[4] In this scenario, the generalized transport mechanism is highly solicited to incorporate the band-to-hopping via intermediate (crossover) transport. With this interest, herein we integrate the flickering resonance-associated degeneracy into the entropy-ruled charge transport theory.[1-3,5,6] The governing equations of differential entropy $(h_S)$ contributions to the energy gap between the adjacent electronic sites $\left(E_g = \Delta E_{ij}\right)$, and relevant electronic coupling ($V_{ij}$) are given by,[6]

$$\Delta E_{ij}(h_S) = \Delta E_{ij} exp\left(-\frac{h_S}{2}\right) \qquad (1)$$

$$\Delta V_{ij}(h_S) = \Delta V_{ij} exp\left(\frac{h_S}{2}\right) \qquad (2)$$

The above equations provide an inverse symmetrical property between the energy gap and electronic coupling (orbital overlapping) between the adjacent states, which is originally derived from the entropy-dependent energy flux principle.[6-8] Here, the coupling or resonance degeneracy value will decrease while the energy gap increases, which is principally incorporated by the factor of $exp\left(\pm\frac{h_S}{2}\right)$. The thermal vibration along with the structural relaxation (geometric reorganization) generally suppresses the electronic coupling via thermodynamic entropy. At the same time, the bias-driven degeneracy increases the electronic coupling and simultaneously reduces the energy gap between the electronic levels, which can be incorporated by differential entropy, $h_S$.[6,7] The quantum features on charge dynamics in different molecular and material media have been generally quantified by this parameter $h_S$.[5,9] In principle, the differential entropy is a logarithmic function of the Gaussian width $(\sigma_G)$ of the carrier wave packet and is defined by $h_S = ln\left(\sqrt{2\pi e}\sigma_G\right)$. Here, the typical transport behaviour, either localization or delocalization dynamics, can be analysed by the Gaussian width. For degenerate systems, the expected $\sigma_G$ value is higher, and hence the differential entropy is increased, which improves the carrier mobility. The delocalized charge transport property is generally expected when the systems are in dynamic disorder. That is, the dynamic disorder is responsible for the degenerate property.[1-4,10,11] In this case, the Gaussian wave packet width of the carrier is relatively larger than that of the static disorder condition, $\sigma_{G,dyn} > \sigma_{G,st}$. In this context, ballistic transport has been anticipated, which is incorporated by the differential entropy-driven transport relations of diffusion-mobility, conductivity, and the molecular Shockley diode principle (see refs.5-7). Therefore, the necessity arises to give more attention to the effect of differential entropy on the electron transfer rate equation, since it is fundamentally connected with all other CT parameters (from the diffusion-mobility to diode principle, via current-voltage characteristics). According to the electronic and nuclear degrees of freedom, the nonadiabatic-to-adiabatic electron transfer shift is anticipated in the biological system, which is directly connected to the flickering resonance degeneracy. With this central attention, the generalized transfer rate equation is here formulated from the entropy (inclusion of degeneracy effect)-accompanied-energy gap/-electronic coupling equation(s) with the Marcus theory,

$$k_{CT} = \frac{2\pi V_{ij}^2 exp(h_S)}{\hbar} \frac{1}{\sqrt{4\pi\lambda k_B T}} exp\left[-\frac{\left(\lambda+(E_j-E_i)exp(-h_S/2)\right)^2}{4\lambda k_B T}\right] \quad (3)$$

Here, the parameters $V_{ij}$ and $\Delta E_{ij} = E_j - E_i$ are computed under static disorder conditions (i.e., zero mode of structural oscillation or fluctuation). With respect to the time scale of structural fluctuation, the existence of degeneracy enrols in to the rate equation, which has a direct proportional to the site-energy matching probability ($P_m$). In this regard, the probability of energy levels crossing/matching ($P_m$) over time provides the electron transfer rate, and the magnitude of $P_m$ is responsible for delocalization transport.[1,3] According to the delocalization/localization character, the carrier distribution (normal Gaussian distribution) width is anticipated, which influences the electronic transport via the exponential factor of differential entropy. The kinetic and potential energy terms are generally correlated with the parameters $V_{ij}$ and $\Delta E_{ij} = E_j - E_i$, respectively. Under static to dynamics disorder transition condition (rigid spacer to nuclear dynamic), the changes in the electronic coupling and site energy difference due to resonance degeneracy can be mapped using Eqns. 1 and 2 as,

$$exp\left(\frac{h_{S,rel}}{2}\right) = \frac{V_{ij,dyn}}{V_{ij,st}} = \frac{\Delta E_{ij,st}}{\Delta E_{ij,dyn}} \rightarrow \left(\frac{\sigma_{G,dyn}}{\sigma_{G,st}}\right)^{\frac{1}{2}} \quad (4)$$

Notably, the resonance degeneracy concurrently enhances the electronic coupling and suppresses the site energy gap in the same order, $V_{ij}(h_{S,f}) - V_{ij}(h_{S,i}) = -\left(\Delta E_{ij}(h_{S,f}) - E_{ij}(h_{S,i})\right) = -\left(E_{g,f} - E_{g,i}\right)$. That is, the degeneracy-weighted coupling and electronic gap can be written as,

$$\delta V_{ij}^* = -\delta E_{ij}^* \quad (5)$$

For dynamic disorder-driven degenerate cases of $V_{ij} \gg \Delta E_{ij}$ and $h_{S,dyn} > h_{S,static}$ (since $\sigma_{G,dyn} \gg \sigma_{G,st}$), the rate Eqn. (3) can be transformed as,

$$k_{CT}^{AD} = \frac{2\pi V_{ij}^2 exp(h_S)}{\hbar} \frac{1}{\sqrt{4\pi\lambda k_B T}} exp\left[-\frac{\lambda - \delta V_{ij}^*}{4k_B T}\right], \qquad (6)$$

since $\Delta E_{ij} exp\left(-\frac{h_S}{2}\right) \to 0$. At this limit, the above Eqn. (6) is suitable for an adiabatic (ballistic) electron transfer mechanism in which $\frac{2V_{ij}exp\left(\frac{h_S}{2}\right)}{\sqrt{2\lambda k_B T}} \geq 1$. This adiabatic limit is hereby mapped using flickering resonance degeneracy along with the entropy-ruled method (see refs. 1, 3 and 6). By implementing Eqns. 1, 2 and 4 into Eqn. 3, the rate becomes,

$$k_{CT} = \frac{2\pi V_{ij}^2}{\hbar}\left(\frac{\sigma_{G,dyn}}{\sigma_{G,st}}\right)\frac{1}{\sqrt{4\pi\lambda k_B T}} exp\left[-\frac{\left(\lambda + \Delta E_{ij}\left(\frac{\sigma_{G,st}}{\sigma_{G,dyn}}\right)^{\frac{1}{2}}\right)^2}{4\lambda k_B T}\right] \qquad (7)$$

At the same time, the nonadiabatic localization transport is anticipated while the electronic coupling is at a too lower value (due to the nondegeneracy effect, $h_S < \frac{S}{k_B}$, where $S$ is the thermal entropy, see refs. 5, 6) and it satisfies the condition of $\frac{2V_{ij}exp\left(\frac{h_S}{2}\right)}{\sqrt{2\lambda k_B T}} \ll 1$. It is here to be noted that the thermal energy contribution dominates over the electronic term. In such a nondegenerate case, the coupling value is reduced and is described by $V_{ij} exp\left(-\frac{S}{2k_B}\right)$.[6] As reported in an earlier study,[3] the charge transfer rate via the flickering resonance method (in terms of dynamic disorder-driven energy levels matching probability) is expressed for 2-site systems as,

$$k_{CT} = \left(\frac{1}{\tau}\right) P_m(2) \qquad (8)$$

where $\frac{1}{\tau} = \frac{\pi V_{ij}(h_S)}{\hbar} \to \frac{\pi V_{ij} exp\left(\frac{h_S}{2}\right)}{\hbar}$, which is related to the Rabi frequency (or inverse of the Rabi time), and $P_m(2) = \frac{2V_{ij}(h_S)}{\sqrt{4\pi\lambda k_B T}} exp\left[-\frac{\left(\lambda + \Delta E_{ij} - V_{ij}\right)^2}{4\lambda k_B T}\right] \cong \frac{2V_{ij}exp\left(\frac{h_S}{2}\right)}{\sqrt{4\pi\lambda k_B T}} exp\left[-\frac{\left(\lambda + \Delta E_{ij} exp\left(-\frac{h_S}{2}\right)\right)^2}{4\lambda k_B T}\right]$; since $\delta E_{ij}^* = -\delta V_{ij}^*$, due to resonance-degeneracy-correlated energy gap quenching, which also simultaneously leads to the resonance-driven electronic coupling (see Eqns. 1, 2, 4, and 5). Therefore, the energy barrier for electron transfer kinetics is subject to the reduceable with respect to the degeneracy-weighted-differential entropy, which facilitates the nonadiabatic to adiabatic transition mechanism, and vice versa too. The generalized charge transfer rate is given by

$$k_{CT} = \left(\frac{1}{\tau^*}\right) exp\left[-\frac{\left(\lambda+\Delta E_{ij}\right)^2}{4\lambda k_B T}\right] \quad (9)$$

Here, $(\tau^*)^{-1} = \frac{2\pi V_{ij}^2}{\hbar}\frac{1}{\sqrt{4\pi\lambda k_B T}}$, which is the general form of an inverse time for charge transfer. Depending upon the resonance-correlated matching probability along with the $h_S$ value, either the $(\tau^*)^{-1}$ acts as a resonance-driven rate, or it belongs to the potential barrier-limited charge trapping rate.[3] This rate equation directly influences the ionic conducting properties (via diffusion) in the bio systems that control many functional activities, from an enzymatic reaction to ionic balancing in organs. The ionic motion is directly proportional to the dynamics of atomic coordinates (e.g., protein folding dynamics) in different time scales, which modifies the active sites' energy profile and hence the functional behaviour too. The complete delocalized adiabatic mechanism (see Eqn. 6) will be achieved when the oscillation (atomic/nuclear motion) frequency is in the order of the harmonic frequency at which the kinetic non-Condon principle will emerge.[4,10,11]

## SIMULATION AND ANALYSIS

To get more insight on ionic (charge) motion along the molecular sites/species, we need to probe the dynamically coupled energetic disorder effect on carrier mobility for both the limited (short range) and extended hopping site (long range) systems. The energetic disorder is defined as the site energy difference between adjacent sites; i.e., the energy difference between the two nearby local minima of the potential energy surface of the reaction coordinate of a system. According to nuclear dynamics, the magnitude of the site energy difference ($\Delta E_{ij}$), matching probability ($P_m$), as well as an interaction strength of coupling ($V_{int} \rightarrow V_{ij}$), are accompanied within the rate equation, and hence the mobility, conductivity, and current are modified. The thermodynamics effect on quantum features on these parameters can be mapped by differential entropy-ruled CT formalism, which connects the resonance correlated degeneracy and transport properties (see refs. 1, 3-6, and 9-12). Here, the site energy difference can be tuned by an external electric field ($\Delta E_{ij}(\vec{E}) = E_j - E_i - q\vec{E}R_{ij}$),[7,11] which influences the charge transfer rate. As observed from the earlier studies (cross references are in),[11,12] the applied field enhances the charge transfer rate in an exponential manner for the dynamical systems of a few hopping sites (limited or short range). For the long-range order of an extended dynamical system (multiple hopping sites), the applied bias field limits the carrier propagation, and hence it leads to charge trapping. This restricted charge transport has been generally observed in the extended or conjugated molecular systems.[12, 13] In general, the charge transport model for an extended system is more suitable for any conjugated/stacked biological systems (DNA, self-assembled molecular units) and other molecular building blocks.[4,10-15] In this line, to map the correlation between geometric fluctuation and energetic disorder for the rate calculation, we reproduce the results from an earlier study,[16] to explore the coherence-to-decoherence charge dynamics, which is plotted in Fig. 1. For this numerical calculation, we choose the extended molecular system of methyl-substituted tris(40-(1-phenyl-1H-benzimidazole-2-yl)biphenyl-4-yl)amine (Me-TIBN) as a test system, and we have performed the Monte-Carlo (MC) simulation to explore the dynamic disorder effect on CT at different sets of $\Delta E_{ij}(\vec{E})$ values of 0, 25. 50, 75, and 100 meV. The CT key parameters such as electronic coupling, site energy difference, and reorganization energy, along with the conformational disorder information from the molecular dynamics (MD) simulation, are taken from Ref. 16.

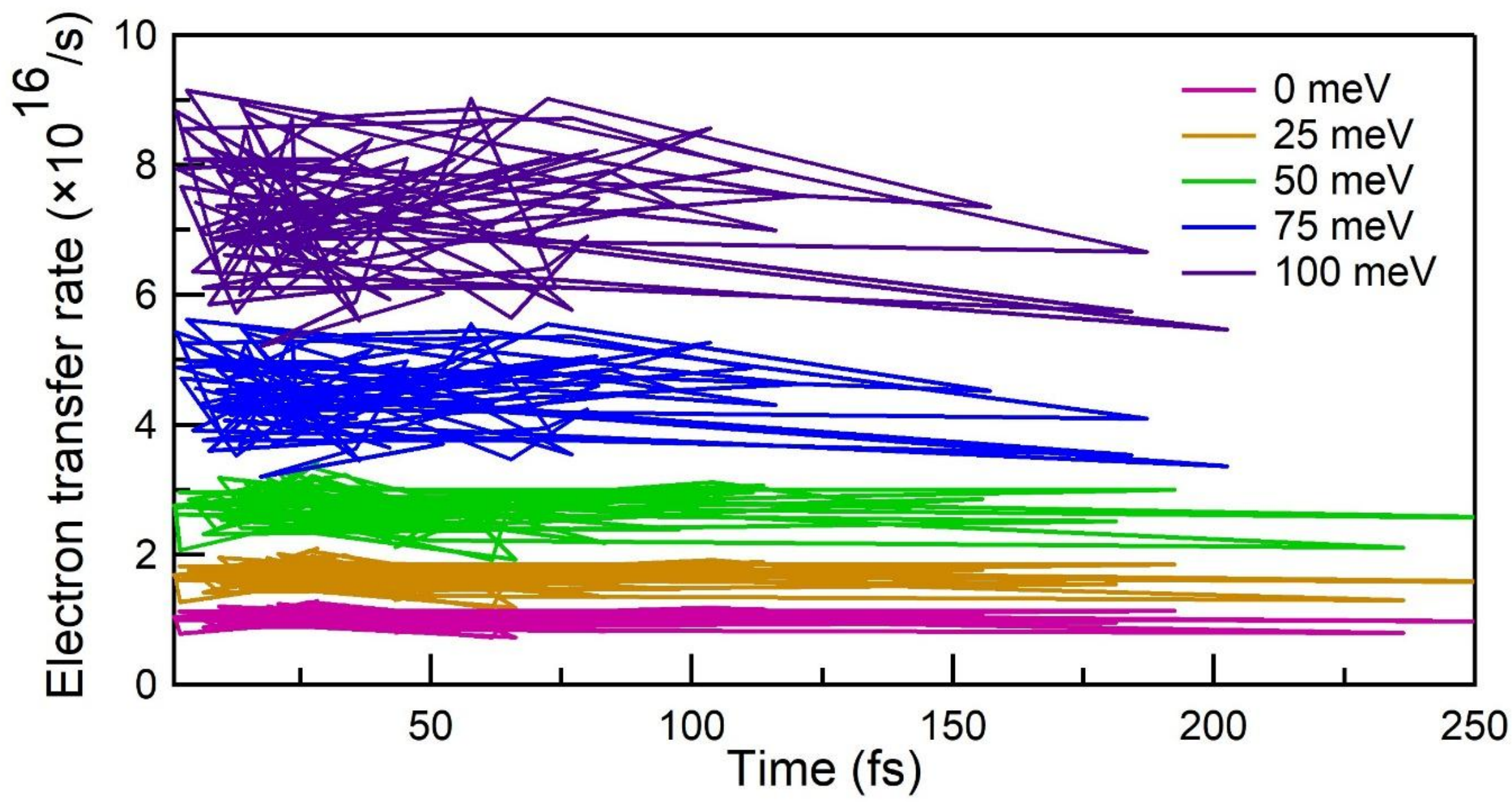

**FIG. 1.** The variation in electron transfer rate at each time step of the simulation is noted under different sets of electric field-assisted site energy difference values of 0, 25, 50, 75, and 100 meV. For the dynamical system of extended sites (long range), the applied bias or electric field drifts the electron transfer rate/kinetics up to a few hopping sites only; after that, it limits the carrier hopping due to the significant loss in rate by energy dispersion at each hopping step, which causes charge trapping. Here, the energy loss at each hopping steps is more for the larger amplitude of site energy fluctuation, which can be analysed by the dispersive parameter (see ref. 11). At the same time, the calculated energy loss is negligible for lower energy flux systems over the extended dynamics systems in which the calculated site energy matching probability is maximum, and hence the coherent-type electron dynamics (ballistic) is anticipated. The transition from dynamic to static disorder is noted, while the site energy gap is increased by the applied electric field at a fixed time scale of nuclear dynamics of a given organic system.

The results show that the applied bias via electric field-assisted $\Delta E_{ij}(\vec{E})$ drives the carrier kinetics up to a few hopping sites; beyond that, the charge is trapped due to dispersion correlated kinetic loss for charge propagation. The energy loss of a carrier at each hopping in the extended systems is strongly depends upon the coupled effect between the structural dynamics on a time scale (oscillation frequency or nuclear dynamics per second) and the amplitude of the applied field via $\Delta E_{ij}(\vec{E})$ parameter. Therefore, the frequency coupled bias voltage or electric field act as a key descriptor for ionic conductivity in an extended dynamical system, like conjugated/stacked bio systems and molecular wires. The dynamics disorder is expected for high oscillation frequency with low energetic disorder (i.e., minimum amplitude of site energy flux, $\Delta E_{ij}(\vec{E})$). Here, the band-like adiabatic transport is anticipated while the oscillation dynamics in the order of harmonic frequency, which is principally connected with the resonance-driven degeneracy. In such that, the dynamically coupled lower bias energy has a minimum kinetic loss in each hopping step, and hence charge propagation takes place over the entire conjugated/stacked length. For instance, the electron carrier is trapped within a few hopping steps for the site energy fluctuation of 100 meV, which is noted in Fig. 1. On the other hand, the loss in charge transfer rate is minimum for zero energy flux of $\Delta E_{ij}(\vec{E})$, which facilitates the electron propagation over the full simulation length (i.e., considered length of a few hundred hopping units). From this investigation, the dynamic to static disorder transport transition is observed when the applied electric field or voltage is increased at a fixed geometric fluctuation (in time scale). In other words, the dynamic disorder reduces as the amplitude of thermal fluctuation reduces or suppresses, which is in good agreement with the earlier report.[17]

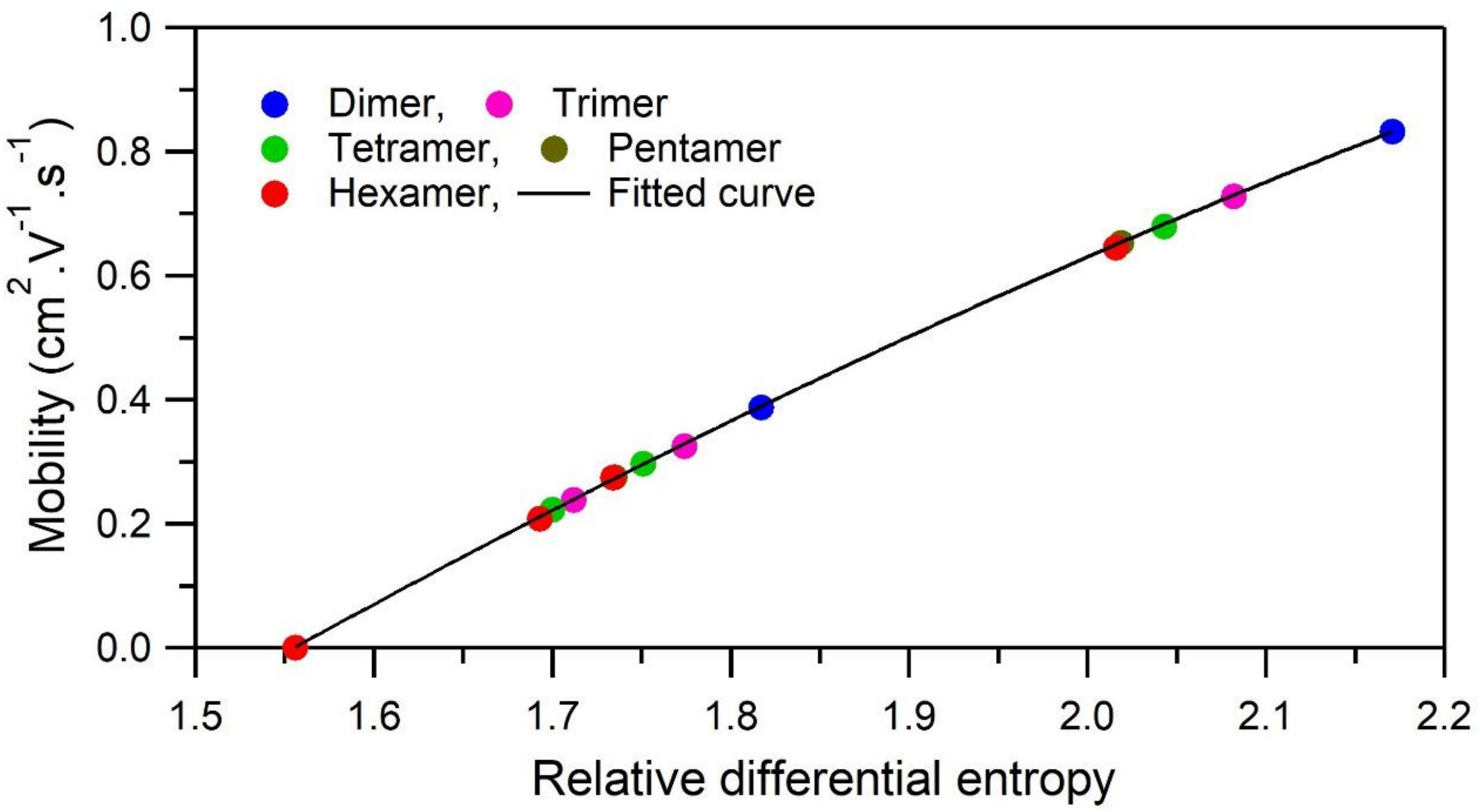

**FIG. 2.** The applied electric field effect (via site energy difference, $\Delta E_{ij}(\vec{E})$) on electron mobility in the different extended dynamical molecular units (from dimer to hexamer) is numerically calculated using the entropy-ruled method for different sets of $\Delta E_{ij}(\vec{E})$ values of 0, 25, 50, and 75 meV. According to the magnitude of the applied field, the existence of degeneracy improves the carrier mobility, which is here quantified through the differential entropy. It is observed that for $\Delta E_{ij}(\vec{E})$ = 75 meV, the calculated drift mobility is maximum for the dimer (blue closed circle) comparable to other extended systems. At the same time, the average mobility of the hexamer (red closed circle) is significantly reduced due to the large carrier energy loss by incoherent-dispersion dynamics. The large applied field suppresses the dynamic disorder, and hence the mobility decreases. The trend of mobility reduction is noted from the right end of a fitted curve (see all closed circles, from blue to red). That is, the mobility decreasing order is relatively more from dimer to hexamer under large energy flux conditions. On the other hand, the mobility decreasing order is much less from dimer to hexamer for small amplitude of $\Delta E_{ij}(\vec{E}) \rightarrow 0$. Here, the dynamic disorder enhances the resonance degeneracy via matching probability, and therefore the mobility decreasing order is minimal due to negligible dispersion, which is noted at the left end of a fitted curve.

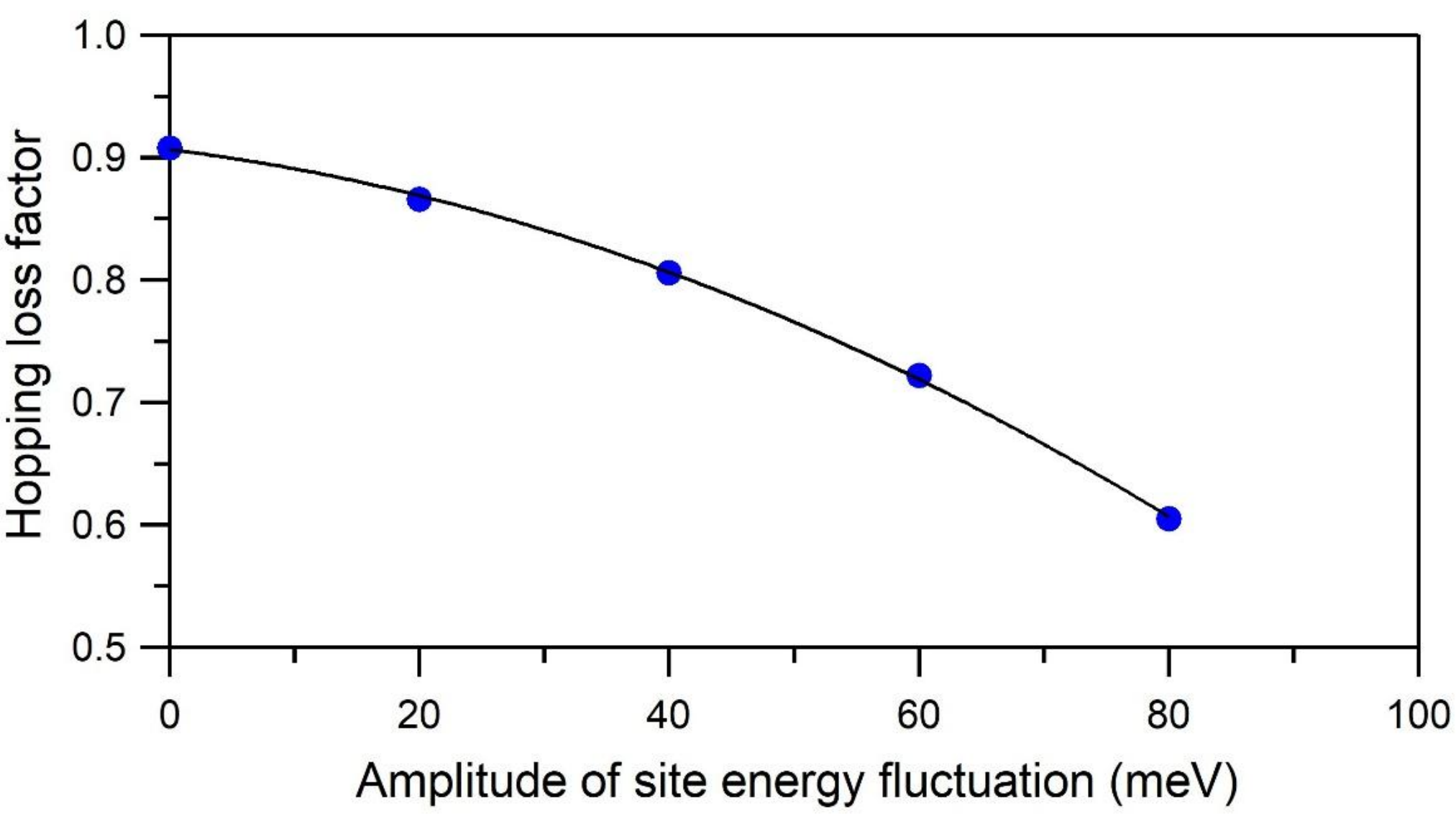

**FIG. 3.** Electron hopping delay is noted while increasing the applied electric field via the site energy difference (0, 20, 40, 60, and 80 meV) along the $\pi$-stacked molecular units of BDHTT-BBT. In this numerical simulation (Monte-Carlo) study, the structural oscillation time is fixed as 1 ps, and the molecular dynamics time step is 1 fs. The carrier propagation state is averaged from every 100 steps of position and velocity coordinates to get a better-approximation. The results are reproduced here from the previous study; see ref. 11 to validate the method. The carrier energy dispersion occurs over the hopping sites in an extended dynamical system, and it increases as the amplitude of $\Delta E_{ij}(\vec{E})$ increases. Here, the dispersion-assisted hopping loss leads to charge trapping in the conjugated polymeric or extended hopping units, which is a more suitable charge transport approach for self-assembled bio aggregates.

In this model, the degeneracy contribution to the charge transport is characterized by differential entropy, which is shown in Fig. 2. Here, the existence of degeneracy levels over the energy (or chemical potential) act as the key descriptor for ionic/charge mobility. At too low electric field condition of $\Delta E_{ij}(\vec{E}) \rightarrow 0$, the calculated mobility values for different extended dynamical units (from dimer to hexamer) almost follows the constant propagation over the entire conjugation length, and hence all closed circles (e.g., blue for dimer, and red for hexamer) overlapped in a single position in the left end of a fitted curve in Fig. 2. On the other hand, there is a quite difference in mobility variation (decreasing from dimer to hexamer) is observed while the applied electric field increases via $\Delta E_{ij}(\vec{E})$, from 0 to 75 meV. This mobility reduction from dimer to hexamer model systems is easily distinguished in the right end (i.e., top region) of a fitted curve in Fig. 2. In this plot, for large $\Delta E_{ij}(\vec{E})$, the mobility difference is clearly apparent for dimer (blue spot), trimer (rose spot), tetramer (green spot), etc. (see, right end of the plot). This reduction in mobility is strongly associated to the geometric fluctuation over the time scale (i.e., oscillation frequency) and amplitude of $\Delta E_{ij}(\vec{E})$, which is normally tuned or controlled by an electric field. Therefore, charge trapping occurs in the extended systems of molecules, especially for bio-aggregates, under strong bias/$\Delta E_{ij}(\vec{E})$ and low oscillation frequency conditions, which causes trap-assisted recombination, which is known as the Shockley-Read-Hall mechanism. At $\Delta E_{ij}(\vec{E})$=25 and 50 meV cases, the mobility reduction (dimer to hexamer) appears between the left and right ends of a fitted curve (around the middle portion of the plot). The present theoretical analysis clearly emphasizes that, under a strong applied electric field, the average rate and mobility of the extended dynamical systems (a few ten hopping sites) have lower values than the limited hopping sites (e.g., a 4-site model), as observed from Fig. 3. Here, the average rate is calculated from the dispersion-weighted equation, $k_{CT}(N) = k_{12}\sqrt[N]{\prod_{N=1}^{N} a_N}$, where $k_{12}$ is the first hopping rate (1st to 2nd site), and $a_N$ is the hopping rate loss factor at each hopping steps, and hence the average mobility also for *N*+1 sites' (or *N*-hopping process) systems is reduced, which is shown in Fig. 3. In other words, an inverse of $\sqrt[N]{\prod_{N=1}^{N} a_N}$ provides the charge transfer time-delay factor. Using a continuum-time-delayed electron hopping method,[12] one can incorporate the short- and long-range ordered transport. Based on the quantum state population, the width of the carrier wave packet will be either narrower or wider; accordingly, the transfer kinetics follows nonadiabatic localization or adiabatic delocalization character, respectively. Therefore, the change in differential entropy over the energy range in the extended (*N* hopping sites/molecular units) biomolecular systems is directly proportional to the density of states (degeneracy population), which is referred to as the DOSP, $\frac{dh_S}{d\eta}$. The mobility increases as DOSP increases, since DOSP times the diffusion constant gives a mobility. The governing relation of the generalized diffusion-mobility relation is given by,[5,6]

$$\left(\frac{\mu}{D}\right)_d = \left(\frac{d}{d+2}\right) q \frac{1}{q}\frac{dh_S}{d\eta} \qquad (10)$$

Where, $d$, $\eta$, and $h_S$ are the dimensions of the carrier network ($d$ = 1, 2, and 3D), chemical potential, and differential entropy $\left(h_S = ln\left(\sqrt{2\pi e}\sigma_G\right)\right)$. For nondegenerate cases, the DOSP is in the order of Franck-Condon-weighted (i.e., thermally populated states) factor, at which the nonadiabatic transition theory is suitable. On the other hand, the adiabatic theory is valid while the DOSP factor is weighted by the degenerate/more interactive electronic states (quantum picture). According to the entropy-ruled method, the charge (or ionic) mobility in biosystems along the consequential electronic sites (here, one-dimensional propagation) can be described as,[5,6]

$$\mu = \left(\frac{1}{5}\right)\frac{q}{k_BT}\frac{D}{\left[\left(\frac{1}{5}\right)+\frac{\left(\lambda+\Delta E(\vec{E})\right)}{2\lambda k_BT}\frac{dE(\vec{E})}{dh_S}\right]} \qquad (11)$$

Here, at finite temperature, the chemical potential for molecular species is defined as $\eta = k_BT\left(1+\frac{1}{3}h_S+\frac{\left(\lambda+\Delta E(\vec{E})\right)^2}{4\lambda k_BT}\right)$. This is the minimum required chemical potential to charge hop over the barrier. It is known that the electrons and ionic charge diffusion continuously takes place in the biosystems, which functionally regulate the metabolic activities and cellular mechanisms. For nonadiabatic transport cases of disordered molecular media, the diffusion has been generally expressed by the Marcus theory of hopping rate equation for a one-dimensional network as

$$\mu = \left(\frac{1}{10}\right)\frac{q}{k_BT}\frac{\langle x^2(t)\rangle k_{ct}}{\left[\left(\frac{1}{5}\right)+\frac{\left(\lambda+\Delta E(\vec{E})\right)}{2\lambda k_BT}\frac{dE(\vec{E})}{dh_S}\right]} \qquad (12)$$

In such a way that the large energy flux between the adjacent active (energy) sites occurs during ionic/electron kinetics in an appropriate channel, which is responsible for non-equilibrium transport too. For the large degeneracy cases (or $T\rightarrow 0$), the above mobility equation is well-approximated as,[7]

$$\mu \approx \left(\frac{2}{5}\right) q \frac{\lambda}{(\lambda+\Delta E_{ij})}\frac{dh_S}{dE(\vec{E})} D \qquad (13)$$

The intermediate CT can be explained by the above mobility equation, and it does not depend on the temperature (see Eqn. 13, absence of thermal activation), since the net ionic charges induce an effective potential, which is too larger than that of thermal energy ($qV_c >> k_BT$). For the adiabatic limit $\left(\Delta E_{ij} = E_j - E_i = 0\right)$, the mobility relation (Eqn. 13) is further reduced to

$$\mu \cong \left(\frac{2}{5}\right) q \frac{dh_S}{dE(\vec{E})} D \qquad (14)$$

This equation is suitable for quasi-quantum systems (i.e., too weak disorder and large delocalized systems), which relates band-like mobility from intermediate range (far from hopping, but close to band transport). In such degenerate and dynamic cases, the energy gap between adjacent sites/residues is further reduced (see Fig. 4); at the same time, the electronic coupling is enhanced in a concurrent manner, and it now obeys Eqns. 1 and 2**.**

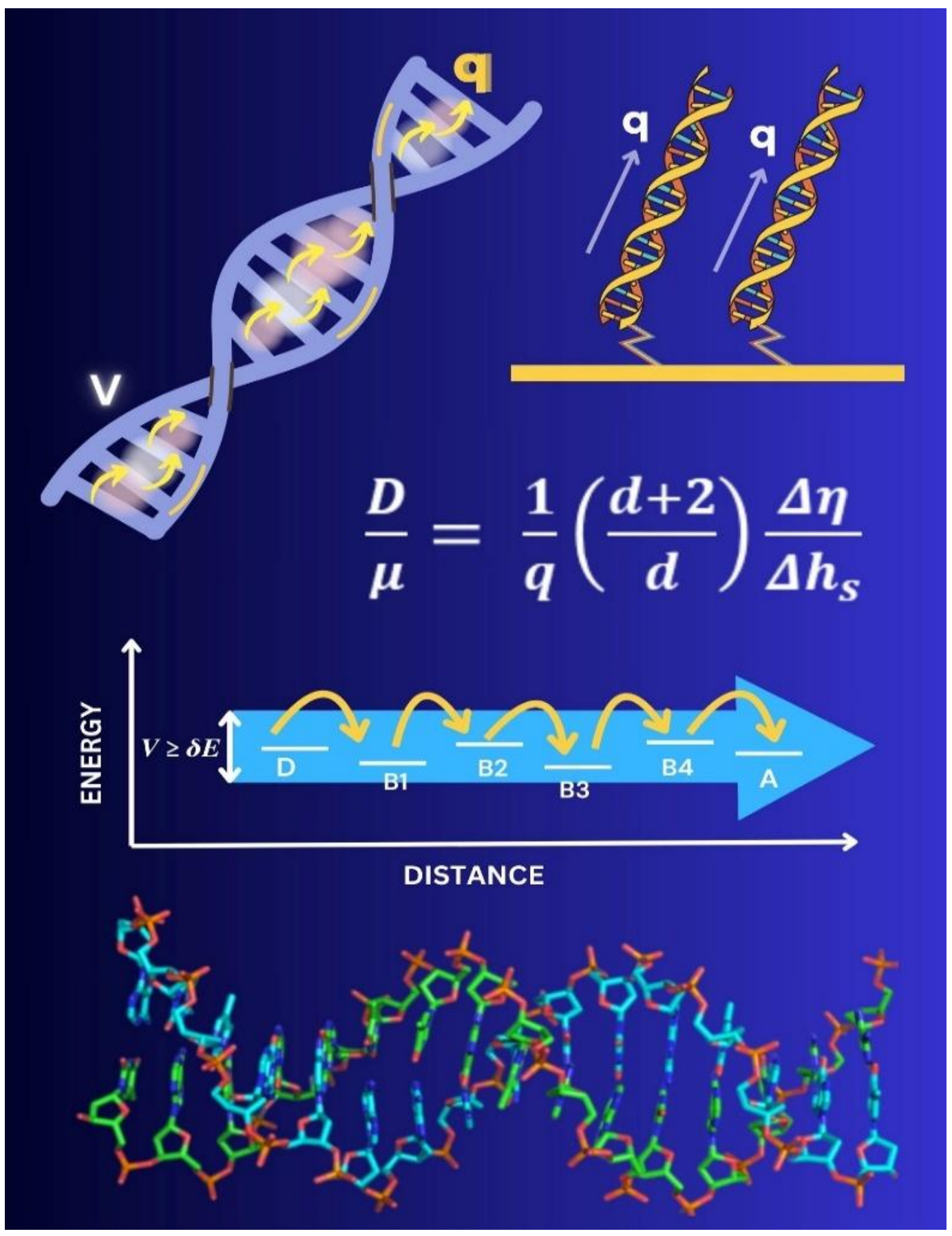

**FIG. 4.** The dynamic disorder-driven flickering resonance effect on charge mobility is incorporated by the entropy-ruled method. The site energy matching probability enhances the degeneracy, which is quantified by differential entropy, $h_S = ln\left(\sqrt{2\pi e}\sigma_G\right)$, where $\sigma_G$ is the Gaussian width. In this relation, $\sigma_G$ is a deterministic parameter to decide the typical transport; whether it is a localization or delocalization transport. The static-to-dynamic disorder transition transport is here generalized using flickering resonance along with the entropy-ruled method,

which describes the transition from nonadiabatic hopping to adiabatic band theory via an intermediate transport regime.

For continuum band states $\Delta h_S = h_{S,f} - h_{S,i} \equiv \frac{\sigma_{G,f}}{\sigma_{G,i}} \rightarrow 1$ (due to large electronic compressibility; see ref. 5, cross references are in), and the kinetic energy of an electron in the conduction region is $E_F = \frac{1}{2} m^* \langle v^2 \rangle$; where $m^*$ and $\langle v^2 \rangle$ are the effective mass of an electron and mean squared velocity, respectively.[18] In such cases, the diffusion coefficient (*D*) will be equal to $\frac{l^2}{2\tau} = \frac{1}{2}\langle v^2 \rangle \tau$. Therefore, the entropy-ruled mobility relation is here transformed for ideal quantum systems to be $\mu = \frac{1}{3} q \frac{\Delta h_S}{\Delta \eta} D \cong q \frac{\Delta h_S}{\Delta E_F} D \rightarrow \frac{q\tau}{m^*}$.[5,6] Recent research reports explain the possibility of Dirac transport properties even in different organic solids at appropriate conditions.[19-21] In this paper, the proposed transport theory is suitable for quantum-classical carrier dynamics in molecules and materials. Here, the energy and diffusion coefficient for the Dirac transport systems are expressed by $E_F = \hbar k v_F$ and $D = \frac{1}{2} v_F^2 \tau$, since the mean free path is $l = v_F \tau$; where $v_F$ and $\tau$ are the Fermi velocity ($v_F$=1×10$^6$ m/s) and relaxation time, respectively. In 1D systems, the change in energy of the systems with respect to the particle flux is related by $\frac{dE}{dN} = \frac{1}{3} E_F$.[5] At low temperature and degenerate limits, the electron transfer rate equation (see Eqn. 6) obeys the adiabatic transition theory and is calculated by $k_{CT}^{AD} = \frac{2\pi V_{ij}^2 exp(h_S)}{\hbar} \frac{1}{\sqrt{4\pi\lambda k_B T}} exp\left[-\frac{\lambda - \delta V_{ij}^*}{4k_B T}\right]$. For nonadiabatic cases, the diffusion coefficient is expressed as, $\frac{\langle x(t)^2 \rangle}{2t}$, for a 1D molecular chain. From this method, the nonadiabatic-to-adiabatic transition is observed, which can be controlled by the parameters of nuclear dynamics' oscillation frequency, amplitude of site energy flux, and importantly resonance degeneracy via the differential entropy.

## IMPLICATIONS FOR PROTEIN PROTECTION FROM REDOX DAMAGE

Proteins arbitrate almost every process that takes place in the cell, exhibiting an almost endless diversity of functions such as signal receptors or transporters to carry specific substances into or out of cells. As the arbiter of molecular function, proteins are the most important final products of information pathways through various functional factors. Among the various mechanisms, the protective activity is essential for stabilizing, in the time scale, in living organisms. Due to enzymatic catalysis and oxidative/reductive stress, the damaging of polypeptides occurs, which diminishes the protective mechanism. For example, the interesting study by Gray and Winkler[22] clearly indicates that human superoxide dismutase mitochondrial

Mn enzyme (SOD2) and mitochondrial cholesterol metabolizing cytochrome P450 (CYP11A1) operate in the $O_2$ processing compartment of the cell, where the risks of reactive oxygen and nitrogen species produced by oxidative stress are high, and hence maximum protection is required.

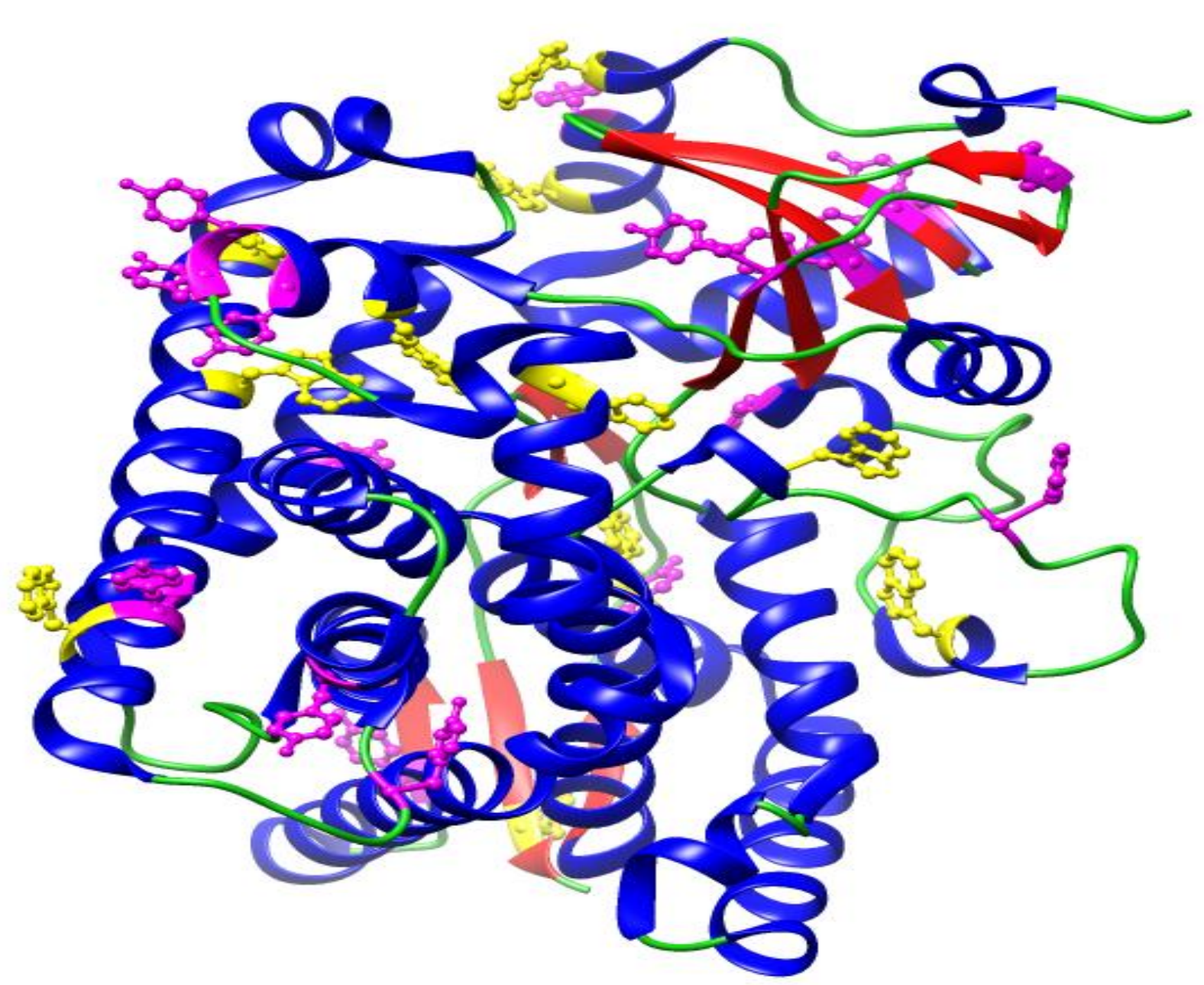

**FIG. 5.** Structural model of human mitochondrial cholesterol-metabolizing cytochrome P450 (CYP11A1, PDB ID: 3N9Y). The tryptophan (Trp) and tyrosine (Tyr) are shown in yellow and purple colours, respectively.

During the reaction, the existence of oxidizing/reduction equivalents (holes/electrons) is transported through the redox-active centres, such as tyrosine and tryptophan residues. Here, the charge transporting ability can enhance the protective mechanism from enzymatic reactions. In this scenario, the location of intermediate states (in the energy profile) is important to study the reactive nature. The position/conformation of the redox-active chains with respect to other units is most important for the formation of intermediate states and electron transfer reactions. The dynamic interaction between the neighboring Tyr/Trp units due to conformational fluctuations modifies the static functional activities and electronic structure effect through appropriate cofactors (such as degeneracy-driven ionic diffusion). Hence, we must turn our view from static to dynamic studies on electron/hole transport and energy transport to understand the real function of metabolism and enzymatic reactions. The model system of interest is SOD2 (PDB ID:1N0J) and CYP11A1 (PDB ID: 3N9Y) due to their metabolic activities. In the previous study of electron transfer kinetics simulations (see ref. 22) of these enzymes, the authors did not include the structural dynamic effect while they studied the survival time for a high-potential intermediate center at the active site. To understand such catalytic functions, one might include the dynamic disorder effect on electron/hole kinetic studies, which we address coalescing the missing information through various co-factors (like degeneracy and conformation disorder effect). This gap was filled by the flickering resonance charge transfer method in biological systems, which was developed by Beratan and his co-workers. However, the cooperative nature of thermodynamics coupled with quantum features on electron and hole carrier transport is highly relevant for dynamic disorder-mediated intermediate formation, diffusion processes, electronic coupling (between the electron donor and acceptor), and site energy fluctuation of adjacent Tyr/Trp residue units, chemical potential gradient, active energy sites oscillation-controlled charge/ionic propagation (leading to accumulation), and flickering resonance. These phenomena can control oxidative damage on protein structure through the fast potential equilibrium, which will destroy the high-energy intermediates and oxidative stress, by the dynamic disorder. On this ground, we introduce a flickering resonance-accompanied entropy-ruled charge transport theory to probe such complex phenomena during various bio functions and metabolic activities, which can provide essential information on carrier network-routed reactive factors and descriptors. Hereby, one can be well-explored any bio functions and enzymatic reactions, including ionic balancing via this transport network-mapped protein's function. The geometric fluctuation amplitude causes the conformational disorder, which typically modifies from a coiled polymer-like structure to a stretched-like geometrical structure; accordingly, the typical transport is anticipated.[23] Based on the nature of conformation as shown in Fig. 5, the intra- and inter-site interactions between different units/residues are increased or decreased, which gives a resultant effect in our charge transfer rate theory. Also, the functional behaviour has been modified by the addition of a foreign agent through medicinal therapy, at which the biomolecular-kinetics (in time scale) associated response nature of active sites is optimized. The dynamic or static fluctuation correlated structure-property (here, charge transport) relationship is closely connected with the thermodynamic entities such as temperature, entropy, and chemical potential. Herein, the geometric dynamics-correlated oxidative stress for protein damage and DNA cleavage with the ionic conduction in different fluid channels can be explored by this flickering resonance incorporated entropy-ruled charge transport theory.

## CONCLUSIONS

In this work, we propose a flickering resonance-accompanied entropy-ruled electron transfer rate theory for limited (a few hopping sites) as well as for extended (multiple sites) model systems, which incorporates both the short- and long-range ordered transport mechanism. The nuclear dynamics-correlated existence of degeneracy is elucidated by the differential entropy, which is closely linked to the sites' matching probability. The thermodynamic effect on the quantum nature of electron dynamics has been theoretically studied for a few sites and an extended system of sites, under different amplitudes of site energy fluctuation. The results show that there is a significant loss in kinetic energy (hopping) of an electron at every hopping step, as further increases with site energy flux. Herein, the thermal vibronic coupled energetic disorder causes the scattered behaviour of hopping, and concurrently, charge trapping takes place within a few hopping steps. It is observed that, for large bias (via electric field or voltage) conditions, the average mobility has a maximum value up to fewer hopping sites only, due to the drift effect by the electric field. Beyond that, the drift effect is controlled by dispersion-associated hopping loss, and hence the charge is trapped. On the other hand, for low-energy bias cases, the carrier drift is not dominant, which results in relatively lower average mobility with minimal hopping loss. In such a case, the charge transport is anticipated for multiple hopping sites (extended long range), but the transfer rate and mobility are comparatively lower than the bias-drift mobility in systems with a few sites. This is a more suitable approach for ionic/proton transport in self-assembled bio aggregates and polymeric systems. Here, a static-to-dynamic disorder transport transition is noted, which can be switched over (from incoherent localization to coherent ballistic) through the amplitude of site energy flux (via applied field) and geometric oscillation frequency at a fixed temperature. Using our transfer rate theory, we address the protein protection mechanism from oxidative stress and some biochemical reactive factors. The position of the intermediate state and reaction barrier can be shifted while the bio-geometric fluctuation transitions from static to dynamic, which is described by resonance accompanied entropy-ruled charge transport theory. The degeneracy-driven differential entropy with time correlation weightage on ionic/charge percolation during enzymatic reactions (externally to fix the reaction rate) is a further highly solicited study to probe any biofunctional activities and ionic balancing in metabolism.

## ACKNOWLEDGMENT

The authors are grateful to the Management of KPR Institute of Engineering and Technology, Coimbatore-641407, India, for providing the necessary computational facility and funds to perform the electronic structure calculations and other conventional quantum mechanics/molecular mechanics studies.

## AUTHOR DECLARATIONS

### Conflict of Interest

The authors have no conflicts to disclose.

### Author Contributions

**Karuppuchamy Navamani:** Conceptualization and Formulation (lead); Problem Designing (lead), Investigation (lead); Discussion and Analysis (equal); Original Draft Preparation (lead); Writing, Review and Editing (equal).

**Muthuvel Pavalamuthu:** Problem Designing (support), Investigation (support); Discussion and Analysis (equal); Original Draft Preparation (support); Writing, Review and Editing (equal).

## DATA AVAILABILITY

The data that support the findings of this study are available from the corresponding author upon reasonable request.